\documentstyle[12pt]{article}
\advance\textheight by \topskip
\begin{document}
\input epsf

\begin{flushright}
ETH-TH/96-43\\
CERN-TH-96\\
DAMTP R/96/48\\
SU-ITP-96-46\\

hep-th/9610155\\
\end{flushright}
\vspace{.5cm}

\begin{center}
\baselineskip=16pt

{\Large\bf  ENHANCEMENT OF SUPERSYMMETRY \\

\

NEAR\, 5d\,  BLACK HOLE HORIZON }

\vskip 1 cm

{\bf Ali Chamseddine,$^a$ ~Sergio Ferrara,$^b$~
Gary W. Gibbons,$^c$~ and \,Renata Kallosh$^d$
}\footnote{E-mail:
%% FOLLOWING LINE CANNOT BE BROKEN BEFORE 80 CHAR
chams@itp.phys.ethz.ch,~~ferraras@cernvm.cern.ch,~~G.W.Gibbons@damtp.cam.ac.uk,\\
\indent ~~kallosh@renata.stanford.edu}

\vskip 1cm

$^a${\em Theoretishe Physik, ETH-H\"onggenberg, CH-8093 Z\'urich,
Switzerland}\\
$^b${\em Theory Division, CERN, 1211 Geneva 23, Switzerland}\\
$^c${\em DAMTP,  Cambridge University, Silver Street, Cambridge CB3 9EW,
United Kingdom}\\
$^d${\em Physics Department, Stanford
University, Stanford, CA 94305-4060, USA}

\end{center}

\vskip 1 cm
\centerline{\bf ABSTRACT}
\vspace{-0.3cm}
\begin{quote}

Geometric Killing spinors which exist on $AdS_{p+2} \times S^{d-p-2}$ sometimes
may be identified with supersymmetric Killing spinors. This explains the
enhancement of unbroken supersymmetry near the $p$-brane horizon in $d$
dimensions. The corresponding $p$-brane interpolates between two maximally
supersymmetric vacua,
at infinity and at the horizon. New case is studied here: $p=0$, $d=5$.
The details of supersymmetric version of the very special geometry are
presented.
We find
the area-entropy formula of the supersymmetric  5d  black holes via the volume
of $S^3$ which  depends on charges and intersection matrix.
\end{quote}

\normalsize
\newpage

\normalsize

\section{Introduction}

The enhancement of supersymmetry near the $p$-brane horizon is an interesting
phenomenon. The corresponding $p$-branes interpolate between two maximally
supersymmetric vacua: ${\cal M}^d$ at infinity and $AdS_{p+2} \times S^{d-p-2}$
near the horizon \cite{G, KP,GT}. The generic reason for the enhancement of
supersymmetry is the following. Any anti-deSitter space as well as any sphere
admit
Killing spinors which we will call geometric Killing spinors. The relation of
this
geometric Killing spinor, which has a dimension of a Dirac spinor,  to the
Killing spinor of unbroken supersymmetry requires
 further investigation. In theories with N=2 supersymmetry the unbroken
supersymmetry
of the $p$-brane
has dimension of  one half of the Dirac spinor. If
near the horizon the dimension of supersymmetric Killing spinor becomes that of
the Dirac spinor, we have a enhancement of supersymmetry.  In some cases it has
already been established that the
Killing spinor defined by the zero mode of the gravitino transformation at the
near horizon geometry of the p-brane solutions
coincides with the geometric spinors. These cases include : d=4, p=0 with
$AdS_{2} \times S^{2}$  \cite{G,KP};
d=10, p=3
with
 $AdS_{5} \times S^{5}$; d=11, p=2  with $AdS_{3} \times S^{7}$ and  d=11, p=5
with
$AdS_{7} \times S^{3}$~
\cite{GT}.  The near near horizon geometry  of these $p$-branes is known to be
maximally supersymmetric. In d=4 the integrability condition for the
Bertotti-Robinson geometry near the black hole horizon  $AdS_{2} \times S^{2}$
was proved in \cite{K, KP} using the fact that this geometry is conformally
flat and that the graviphoton field strength is covariantly constant.

Remarkably, the supersymmetry near the 5d black hole horizon has not yet been
studied. Moreover,  the unbroken supersymmetry of the 5d black holes was
established in \cite{GKLTT} only for black holes of pure N=2, d=5 supergravity
without vector multiplets. In particular the Strominger-Vafa 5d black holes
\cite{SV} and the rotating generalization of them  \cite{rot} have not yet been
embedded into a particular 5d supersymmetric
theory and the unbroken supersymmetry of either solution has not been
checked directly. An analogous situation holds with the more  general 5d static
and rotating black
holes  found in \cite{TseyCvY}. There are various indications,
however,  that these solutions have unbroken supersymmetry.

 The purpose of this paper is to find out  whether the 5-dimensional black
holes near the horizon show the enhancement of the supersymmetry.  Specifically
we will try to identify the supersymmetric Killing spinors admitted by black
holes near the horizon with the geometric Killing spinors of  $AdS_{2} \times
S^{3}$. We will also derive the area formula for generic solutions in N=2
theory interacting with arbitrary number of vector multiplets as the volume of
the $S^3$ and express it as the function of charges and the intersection
matrix.

For the
$d$-dimensional manifold which is a product space $AdS_{p+2} \times S^{d-p-2}$
the geometric Killing spinors are given by the product  of Killing spinors on
$AdS_{p+2} $ and $ S^{d-p-2}$.
On  $AdS_{p+2} $ and on $S^{d- p-2} $ the Killing spinor equations are
\begin{eqnarray}
\hat \nabla_{a} \eta (x) &\equiv& ( \nabla_{a} + c_1\tilde \gamma_{a })  \eta
(x)=0\ ,
\qquad a =0,1,\dots , p+1 \ , \\
\nonumber\\
\hat \nabla_{\alpha } \varsigma (y) &\equiv& ( \nabla_{\alpha} +
c_2\gamma_{\alpha} )  \varsigma (y)=0\ ,  \qquad \alpha =p+2, \dots , d-1.
\end{eqnarray}
Here $\gamma_\alpha$ is the $\gamma$-matrix of the $d-p-2$-dimensional
Euclidean space and the commutator $[\tilde\gamma_a,  \tilde \gamma_b]$ equals
$- [ \gamma_a,  \gamma_b]$ where
$\gamma_a$ is the $\gamma$-matrix of the $p+2$-dimensional Minkowski space.

The integrability conditions for the geometric Killing spinors defined above is
\begin{eqnarray}
&&[ \hat \nabla_{a}, \hat \nabla_{b} ] \eta (x) = 0 \qquad \Longrightarrow
\qquad R_{ab}{}^{cd} = 4  ( c_1)^2 (\delta_{a}{} ^{c}
\delta_{b}{} ^{d} - \delta_{a}{} ^{d}
\delta_{b}{} ^{c})\\
\nonumber\\
&&[ \hat \nabla_{\alpha}, \hat \nabla_{\beta} ] \varsigma (y) = 0  \qquad
\Longrightarrow  \qquad R_{\alpha\beta}{}^{\gamma \delta} = -4 (c_2)^2(
\delta_{\alpha}{} ^{\gamma} \delta_{\beta}{} ^{\delta} -
\delta_{\alpha}{} ^{\delta} \delta_{\beta}{} ^{\gamma})
\end{eqnarray}
and it is satisfied by the geometries of the Anti-deSitter space and sphere.
The full Killing spinor is
\begin{equation}
\epsilon (x,y) = \eta(x) \varsigma (y)
\end{equation}
and it forms a Dirac spinor in a given dimension $d$. On the other hand the
Killing spinor of unbroken supersymmetry of the $p$-brane solution near the
horizon is defined by the corresponding supersymmetry transformation of the
gravitino. If  supersymmetric Killing spinor near the horizon can be identified
with the geometric one of the maximally supersymmetric product space,  we have
an enhancement of supersymmetry:
\begin{equation}
\delta \Psi_{M} =0 \qquad \Longrightarrow \qquad \hat \nabla _{M} \epsilon
(x,y) = 0    \Longrightarrow \left (\matrix{
\hat \nabla _{\mu} \eta (x) =0\cr
\hat \nabla _{\alpha }\varsigma (y) =0\cr
}\right )\ .
\end{equation}

To study this issue we will work out some details of d=5 N=2 supergravity
coupled to N=2 vector multiplets  in the framework of very special
geometry, see
Sec. 2. This formulation of the d=5 theory is particularly adapted to d=11
supergravity compactified on Calabi-Yau manifold characterized by the
intersection number $C_{ABC}$. In Sec. 3 we will identify the supersymmetric
Killing spinors admitted by black holes near the horizon with the geometric
Killing spinors of  $AdS_{2} \times S^{3}$. We will also perform  detailed
derivation of the area formula of the five-dimensional  black holes \cite{FK1}
in terms of
the volume of $S^3$ which we evaluate in terms of the charges and intersection
number matrix.
In Sec. 4 we will consider a special case of the N=2 theory
interacting with one vector multiplet, which comes from the truncation of N=4,
d=5 supergravity. This will allow us to embed  some of the known black
holes into supersymmetric theories and study the enhancement of supersymmetry
near the horizon. We also  study the N=4 theory  interacting with arbitrary
number $n$ of N=4 vector multiplets as a particular example of N=2 theory
interacting with $n+1$
N=2 vector multiplets and intersection matrix $C_{oij}= \eta_{ij}$, where
$\eta_{ij}$ is the Lorenzian metric on $ (1,n)$. The  duality invariant
area-entropy formula for the  N=4 theory is truncated to N=2 theory as it was
done before in  four-dimensional theory  in \cite{FK2}. In this way one
describes the theory with  very special geometry
$ O(1,1)\times O(1,n)/O(n)$.
In the discussion section we explain the relation between the
enhancement of supersymmetries and finiteness of the area of the horizon and
speculate about other
cases as yet not known.

\section {Supersymmetric very special geometry in d=5}
The action for d=5, N=2 supergravity coupled to  N=2 vector multiplets has been
constructed by  G\"unaydin,  Sierra  and  Townsend \cite{GST}. The bosonic part
of the action has been  adopted to the very special geometry \cite{dWvP} and
the compactification of  11d supergravity down  to five dimensions
 on Calabi-Yau 3-folds
\cite{Cadavid} with Hodge numbers ($h_{(1,1)}, h_{(2,1)})$ and
topological intersection form $C_{IJK}$. For our purpose, it will be
extremely useful to adopt the full action and the supersymmetry transformation
laws of \cite{GST} to that of very special geometry.  In what follows we will
give the detailed
derivation of some formulae reported in \cite{FK1}.

The fields of the theory are:
$e_\mu{}^m,~A_\mu{}^I,~\phi^i,~\psi_{\mu r},~\lambda_{r}{}^i$ where
$I=0,1,\dots , h_{(1,1)}-1$,  ~$i=1,\dots$,~$h_{(1,1)}-1$, and $r=1,2$. The
index
$r$ on the spinors is raised and lowevered with the symplectic metric $\epsilon
_{rs}$ and will be omitted.

The  N=2 d=5 supersymmetric Lagrangian describing the coupling of vector
multiplets to supergravity is determined by one function which is given by the
intersection form on a CY 3-fold:
\begin{equation}
{\cal V} = {1\over 6} C_{IJK} X^I X^J X^K \ .
\label{prep}\end{equation}

The action is
\begin{eqnarray}
e^{-1} {\cal L} &=& -{1\over 2} R  -{1\over 2} \bar \psi_\mu
\Gamma^{\mu\nu\rho} D_\nu \psi_\rho - {1\over 4} G_{IJ} F_{\mu\nu} {}^I
F^{\mu\nu J}
-{i\over 2}\bar \lambda^i (g_{ij} \Gamma^\mu  D_\mu + \Gamma^\mu  \partial_\mu
\phi^k \Gamma_{kj}^l
g_{li})\lambda^j
\nonumber\\
&-&{1\over 2} g_{ij} \partial_{\mu} \phi^i \partial^\mu \phi^j
-{i\over 2}\bar \lambda_i  \Gamma^\mu \Gamma^\nu \partial _\nu \phi^i +
{1\over 4} \left({3\over 4}\right)^{2/3} t_{I,i} \bar \lambda^i  \Gamma^\mu
\Gamma^{\lambda \rho} \psi_\mu  F_{\lambda \rho}^I\nonumber\\
&+&{i\over 16\cdot 6^{1/3}} \Bigl(g_{ij} t_I - 9 C_{JKL} t^J_{,i} t^K_{,j}
t^L_{,k} t_I^{,k}\Bigr)-
 {3 i\over 16\cdot 6^{1/3}} t_I \Bigl(  \bar \psi_\mu \Gamma^{\mu\nu\rho\sigma
} \psi_\nu F_{\rho \sigma}^I + 2  \bar \psi^\mu  \psi^\nu F_{\mu\nu}^I\Bigr)
\nonumber\\
&+&{ e^{-1}  \over 48} \epsilon^{\mu\nu\rho\sigma\lambda} C_{IJK} F_{\mu\nu}^I
F_{\rho\sigma}^J A_\lambda^K + \dots \ ,
\end{eqnarray}
where we use  dots  for 4-fermionic terms. Here $t^I= t^I(\phi)$ are the
special
coordinates subject to the conditions
\begin{equation}
t^I t_I=1 , \qquad C_{IJK} t^I t^J t^K =1
\label{special}\end{equation}
and $t_I$ are the ``dual coordinates"
\begin{equation}
t_I = C_{IJK}  t^J t^K \equiv C_{IJ} t^J \ ,\qquad  t^I = C^{IJ}t_J \ .
\end{equation}
We have introduced a notation
\begin{equation}
C_{IJ} \equiv C_{IJK}t^K \ , \qquad C^{IJ} C_{JK} = \delta^I{}_K \ .
\end{equation}
The metric is derived from the prepotential (\ref{prep}) through the relation
($\partial_I \equiv {\partial \over \partial X^I})$
\begin{equation}
G_{IJ} = -{1\over 2} \partial_I \partial_J (\ln {\cal V})|_{{\cal V} =1}
\end{equation}
and $X^I$ is related to $t^I$ by
\begin{equation}
X^I = 6^{1/3} t^I |_{{\cal V} =1} \ .
\end{equation}
Finally, the metric $g_{ij}$ is given by
\begin{equation}
g_{ij}=  G_{IJ} X^J{}_{,i} X^K{}_{,j}=  -3 C_{IJK} t^I t^J{}_{,i} t^K{}_{,j} \
{}.
\end{equation}
We can express $G_{IJ}$ in terms of $t^I$ through the equations
\begin{equation}
G_{IJ} = -{ 6^{1/3} \over 2} (C_{IJ} - {3\over 2} t_I t_J) \
\end{equation}
and the inverse metric acting on vectors is
\begin{equation}
 G^{IJ} = -{2\over 6^{1/3}} (C^{IJ}- 3 t^I t^J)   \ , \qquad G^{IK} G_{KJ}=
\delta^I{}_J\ .
\end{equation}
We will need in what follows the relation between the derivative of the special
coordinate $(t^I)_{,i}$ and that of the dual coordinate
$t_{I,i}$. Using the fact that $C_{IJK}$ is a symmetric numerical tensor we
have
\begin{equation}
t_{I,i} = 2 C_{IJK} (t^J)_{, i} \,  t^K = 2 C_{IJ} (t^J)_{, i} \ , \qquad
(t^J)_{, i} =
{1\over 2} C^{IJ} t_{I,i}\ .
\label{dif1}\end{equation}
Differentiating eqs. (\ref{special}) we get
\begin{equation}
 C_{IJK} t^I t^J (t^K)_{,i} =0\ ,  \qquad  t_I (t^I)_{,i} = t_{I,i} t^I=0 \ .
\label{dif2}\end{equation}
The supersymmetry transformation laws are:
\begin{eqnarray}
\delta e_\mu{}^m &=& {1\over 2} \bar \epsilon \Gamma^m \psi_\mu \ ,\nonumber\\
\delta\psi_\mu &=& D_\mu (\hat \omega) \epsilon +  { i\over 8 \cdot 6^{1/3}}
t_I
\Bigl(\Gamma_\mu{}^{\nu\rho} - 4 \delta_\mu{}^ \nu \Gamma^\rho\Bigr) \hat
F_{\mu\nu}{}^I \epsilon  \nonumber\\
&+&{1\over 12} g_{ij} \left({1\over 4} \Gamma_{\mu\nu\rho} \epsilon \bar
\lambda^i \Gamma^{\nu\rho}\lambda^j - \Gamma_{\mu\nu} \epsilon \bar \lambda^i
\Gamma^{\nu}\lambda^j
- \Gamma^\nu \epsilon \bar \lambda^i \Gamma_{\mu\nu}\lambda^j +2 \epsilon
\bar \lambda^i \Gamma_{\mu}\lambda^j \right)\ ,\nonumber\\
\delta A_\mu ^I &=& 6^{1/3} \left( {1\over 2} t^I{}_{,i} \bar \epsilon
\Gamma_\mu \lambda^i + i \bar \psi \epsilon t^I\right)\ , \nonumber\\
\delta \lambda _i &=& \delta\phi^k\, \Gamma_{ki}{}^l\, \lambda_l + {1\over 4}
\left({3\over 4}\right)^{2/3} t_{I,i}   \Gamma^{\mu\nu} \epsilon   F_{\mu\nu}^I
- {i\over 2} g_{ij} \Gamma^\mu  \partial_\mu  \phi^j \epsilon \nonumber \\
&+& {3i\over 16} t^I{}_{,i}\, t^J{}_{,j}\, t^K{}_{,k}\, \Bigl(-3 \epsilon \bar
\lambda ^j \lambda^k +
\Gamma^\mu \epsilon \bar \lambda ^j  \Gamma^\mu \lambda^k+{1\over 2}
\Gamma_{\mu\nu} \epsilon \bar \lambda ^j \Gamma^{\mu\nu}
\lambda^k\Bigr) \ , \nonumber\\
\delta\phi^i &=& {i\over 2} \bar \epsilon \lambda^i \ .
\end{eqnarray}

\section{Near horizon geometry and supersymmetry}
Double-extreme black holes  (which have 1/2 of unbroken supersymmetry and have
constant moduli \cite{KSW})  in five dimension have the geometry of  the
extreme  Tangherlini solution \cite{Tangh}.  It is a 5d analog of the extreme
Reissner-Nordstrom metric.
\begin{equation}
ds^2 =- \left (1-({r_0\over r} )^2 \right)^2 dt^2 + \left (1-({r_0\over r} )^2
\right)^{-2} dr^2 + r^2 d\Omega_3^2
\end{equation}
and
\begin{equation}
2 \sqrt {- g} \; G_{IJ} F_{tr}^J = q_I \ ,  \qquad \phi^i={\rm const} \ .
\end{equation}
The horizon is at $r=r_0$ where the parameter $r_0$ defining the horizon as
well as the constant values of moduli depend   on charges of the vector fields
and on  the topological
intersection form $C_{IJK}$. Our study of the near horizon geometry will
allow us to determine this dependence.

Note that the area of the horizon of the black hole is given by the volume of
the 3-dimensional sphere
$$A = 2\pi^2 r_0^3 \ .$$
The area formula of 5d black holes was found in \cite{FK1} from the observation
that the unbroken supersymmetry near the black hole horizon requires that the
central charge $Z\sim t^Iq_I$ has to be extremized in the moduli space, i.e.
near the horizon
$\partial_i Z=0$. This leads to the area formula in the form $A\sim (q_I q_J
C^{IJ}|_{\partial_i Z=0}  )^{3/4}$, where $C^{IJ}$ is the inverse of $C_{IJ}=
C_{IJK}t^K$.  In the derivation of this area formula it was assumed that the
unbroken supersymmetry of the black hole solution is enhanced near the horizon.
This will be proved now.
 It will be also explained why  the enhancement
of supersymmetry near the horizon requires the extremization of the central
charge
for describing the area formula.
By exactly solving the Killing equations
for the near horizon geometry we will be able to justify the area formula
suggested in \cite{FK1} and find the explicit area formula including the
numerical factor in front of it.

Near the horizon at $r\rightarrow r_o$  and one can exhibit the $AdS_{2} \times
S^{3}$ geometry using  $\hat r = (r-r_0 )\rightarrow 0$
\begin{equation}
ds^2 =- ({ 2\hat r \over r_0})^2 dt^2 + ({ 2\hat r \over r_0})^{-2}   d \hat
r^2 + r_0^2 d\Omega_3^2 \ .
\label{BR}\end{equation}
Since we deal with the product space we may  use $a={0,1 }$ for the coordinates
of the $AdS_2$-space and $\alpha =2,3,4$ for the coordinates of the 3-sphere
(in tangent space). The vector field ansatz near the horizon becomes
\begin{equation}
2  (r_0)^3 \; G_{IJ} F_{ab}^J = \epsilon_{ab} q_I \ .
\label{vec}\end{equation}
Let us use this ansatz in the fermionic part of supersymmetry transformations
with all vanishing fermions and constant moduli. We start with gaugino and keep
only relevant terms
\begin{equation}
\delta \lambda _i =  {1\over 4} \left({3\over 4}\right)^{2/3} t_{I,i}
\Gamma^{\mu\nu} \epsilon   F_{\mu\nu}^I =0 \ .
\end{equation}
We study the possibility that the zero mode of this equation is given by the
full size spinor $\epsilon$ without linear constraints on it, i.e. that the
unbroken supersymmetry is indeed enhanced near the horizon. This is possible,
provided that
\begin{equation}
t_{I,i}   G^{IJ} q_J=0 \ .
\label{gaug}\end{equation}

 The enhancement of unbroken supersymmetry
near the horizon which can be  deduced from the gaugino part of supersymmetry,
can be represented
as the condition of the minimization of the central charge, as found in
\cite{FK1}.
Let us derive this here in a more detailed way. The gravitino transformation
\begin{equation}
\delta\psi_\mu = D_\mu ( \omega) \epsilon +  { i\over 8 \cdot 6^{1/3}} t_I
\Bigl(\Gamma_\mu{}^{\nu\rho} - 4 \delta_\mu{}^ \nu \Gamma^\rho\Bigr)
F_{\mu\nu}{}^I \epsilon =0
\label{gravitino}\end{equation}
shows that  the graviphoton  field strength is given by the linear combination
of vector fields and moduli  $t_I  F_{\mu\nu}{}^I$,  and therefore the central
charge is proportional to $t^I q_I$.
{}From eq. (\ref{gaug}) we get
\begin{equation}
t_{I,i}   (C^{IJ}- 3 t^I t^J) q_J=0 \ .
\label{gaug2}\end{equation}
Using eqs. (\ref{dif1}) and  (\ref{dif2}) we can conclude that
\begin{equation}
t_{I,i}   (C^{IJ}- 3 t^I t^J) q_J= 2 (t^J)_{,i} q_J =0 \qquad \Longrightarrow
\qquad \partial_i Z=0 \ .
\label{gaug3}\end{equation}
Thus we have derived the condition of minimization of the central charge
$Z=t^I q_I$ in the moduli space from the requirement that the gaugino
supersymmetry transformation for constant moduli has the  full size spinor
$\epsilon$ as a zero mode, i.e. from the condition of enhancement of
supersymmetry near the black hole horizon.  The central charge has to be
independent of $\phi^i$. Let us consider some useful identities of the real
special geometry which are valid only near horizon. Note that
\begin{equation}
g^{ij} \partial_i Z \partial_j Z = g^{ij} t^I{}_{,i} t^J{}_{,j} \,q_I q_J
\equiv \Pi ^{IJ} q_I q_J =0 \ .
\label{identity1}\end{equation}
Using eq. (\ref{dif2}) we find that
\begin{equation}
\Pi ^{IJ} t_I  =0
\end{equation}
and we may look for the combination which is orthogonal to $t_I$  in the form
$\Pi^{IJ}= l (C^{IJ}-  t^I t^J) $. To get the coefficient $l$ we use $g_{ij} =
6^{2/3}
t^I{}_{,i} t^J{}_{,j} G_{IJ}$ and contract it with $g^{ij}$
\begin{equation}
g_{ij} g^{ij} = h_{(1,1)}-1 = 6^{2/3} \Pi ^{IJ} G_{IJ} \ ,
\end{equation}
which leads to $l= -{1\over 3}$. Thus we conclude that near the horizon where
the central charge is moduli independent we have an identity
\begin{equation}
\left ((C^{IJ}  -  t^I t^J)q_I q_J \right)_{\partial_i Z=0} =0 \ .
\label{identity2}\end{equation}

What remains to be done to make the extremization of the central charge
consistent is to check what happens with
the gravitino: does the gravitino transformation rule admit the Killing spinor
of the full supersymmetry in our background?  And what are the conditions of
that?
Using the ansatz for the vector fields near the horizon (\ref{vec}) and taking
into account that we have a product space we get  the following form of the
gravitino transformations (\ref{gravitino}).
\begin{eqnarray}
\delta\psi_a= (\hat \nabla_{a} )\eta  &\equiv& ( \nabla_{a} +  c \tilde
\gamma_{a })  \eta =0\ ,
\qquad a =0,1,
\nonumber\\
\delta\psi_\alpha = (\hat \nabla_{\alpha }) \varsigma  &\equiv& (
\nabla_{\alpha} -{c\over 2} \gamma_{\alpha} )  \varsigma =0\ ,  \qquad \alpha
=2,3,4,
\end{eqnarray}
where
\begin{equation}
\tilde \gamma_a = i \epsilon_{ab} \gamma^b \ , \qquad \Gamma^a = \gamma^a
\otimes I \ , \qquad \Gamma_\alpha = i \gamma_0 \gamma_1 \otimes \gamma_\alpha\
,
\end{equation}
and
\begin{equation}
c= - {1 \over 6^{2/3}} {t^Iq_I \over r_0^3} \ ,
\end{equation}
and the $\gamma$-matrices satisfy
\begin{eqnarray}
\{\tilde\gamma_a, \tilde\gamma_b\} &=& 2\eta_{ab} \ , \qquad \eta_{ab} =(
-,+) \ , \qquad  \tilde \gamma_{[ab]} = - \gamma_{[ab]} \ ,\\
\{\gamma_\alpha , \gamma_\beta \} &=& 2\delta_{ab} \ ,  \qquad \alpha=2,3,4.
\end{eqnarray}

This is a special example of the general case presented in the introduction.
We have identified the supersymmetric Killing spinor with the geometric Killing
spinor.
The integrability conditions for the geometric Killing spinors defined above
are
\begin{eqnarray}
&&[ \hat \nabla_{a}, \hat \nabla_{b} ] \eta (x) = 0 \qquad \Longrightarrow
\qquad R_{ab}{}^{cd} = {4\over 6^{4/3}} {(t^Iq_I)^2 \over r_0^6} (\delta_{a}{}
^{c}
\delta_{b}{} ^{d} - \delta_{a}{} ^{d}
\delta_{b}{} ^{c}) \ .\\
\nonumber\\
&&[ \hat \nabla_{\alpha}, \hat \nabla_{\beta} ] \varsigma (y) = 0  \qquad
\Longrightarrow  \qquad R_{\alpha\beta}{}^{\gamma \delta} = - {1\over 6^{4/3}}
{(t^Iq_I)^2 \over r_0^6}  (
\delta_{\alpha}{} ^{\gamma} \delta_{\beta}{} ^{\delta} -
\delta_{\alpha}{} ^{\delta} \delta_{\beta}{} ^{\gamma})\ .
\end{eqnarray}
This can be contracted to give us the Ricci tensors on $AdS_{2}$ and on
$S^{3}$:
\begin{equation}
R_{ab} =  {4\over 6^{4/3}} {(t^Iq_I)^2 \over r_0^6} \eta_{ab} \ ,  \qquad
R_{\alpha \beta} =
 - {2\over 6^{4/3}} {(t^Iq_I)^2 \over r_0^6} \delta_{\alpha \beta}\ ,
\label{integr}\end{equation}
with the result that the radii of $AdS_{2}$ and  $S^{3}$ are related:
\begin{equation}
R_{(2)} = -{4\over 3}\,  R^{(3)}\ .
\end{equation}
This relation restricts the properties of geometric Killing spinors. They
would exist  without any relation between these two product geometries.
However, supersymmetric Killing spinors require  relation between geometries.
The curvature can be calculated also directly from the metric (\ref{BR}):
 \begin{equation}
R_{ab} =  {4 \over r_0^2} \eta_{ab} \ ,  \qquad R_{\alpha \beta} =
 -  {2 \over r_0^2} \delta_{\alpha \beta}\ .
\label{curv}\end{equation}
Comparing eqs. (\ref{integr}) with eqs. (\ref{curv}) we can express $r_0$ via
the values of the moduli near the horizon and electric charges as follows:
\begin{equation}
r_0^4 = {1\over 6^{4/3}} \left ( (t^Iq_I)_{\partial i Z=0} \right)^2\ .
\end{equation}
This gives us the area of the horizon
\begin{equation}
A = 2\pi^2 r_0^3 = {\pi^2\over 3} \left\{ (t^I t^J q_I q_J)_{\partial i
Z=0}\right\}^{3/4} \ .
\end{equation}

Near the horizon at $\partial i
Z=0$ we  can rewrite it as
\begin{equation}
A =    {\pi^2\over 3} \left\{ (C^{IJ} q_I q_J)_{\partial i
Z=0}\right\}^{3/4}
\end{equation}
using identity (\ref{identity2}).

Thus we have combined the supersymmetry analysis with the analysis of the
geometry. In this way we have  confirmed the structure of the area formula of
five dimensional  black holes in N=2 theories obtained in \cite{FK1}
and found the exact numerical coefficient in front of it for supersymmetric
five dimensional
black holes with finite area of the horizon in the N=2 theory.

\section{Truncation of N=4 supergravity  to N=2}
In some cases it is useful   to consider those N=2 theories which can be
obtained by truncation from N=4 theories. We will first study the truncation of
pure N=4 supergravity and later generalize the result  to the case of  N=4
supergravity interacting with arbitrary number of vector multiplets as it was
done before in 4d theories in \cite{FK2}.

We focus on a a special example of the double extreme five dimensional black
hole
known as Strominger-Vafa black hole \cite{SV}. The  Lagrangian which allows a
supersymmetric embedding of this black hole is obtained most easily by the
truncation of
N=4 supergravity in d=5 constructed by Awada and Townsend \cite{AwTow}. The
bosonic part of the action is
\begin{eqnarray}
e^{-1} {\cal L} &=& -{1\over 2} R   - {1\over 4} e^{{2\over 3}\phi} F_{\mu\nu}
{}^{ij}F_{ij}{} ^{\mu\nu } - {1\over 4} e^{-{4\over 3}\phi} G_{\mu\nu} G
^{\mu\nu }
\nonumber\\
&-&{1\over 6} ( \partial_{\mu} \phi) ^2   +{ e^{-1}   \over 4 \sqrt2}
\epsilon^{\mu\nu\rho\sigma\lambda} F_{\mu\nu}{}^{ij}
F_{\rho\sigma ij } B_\lambda \ ,
\end{eqnarray}
where $i,j = 1,\dots ,4$ and $G_{\mu\nu}$ is the field strength of $B_\mu$ and
$F_{\mu\nu} {}^{ij}$is the field strength of $A_\mu{}^{ij}$.
The supersymmetry transformation laws are:
\begin{eqnarray}
\delta e_\mu{}^m &=& {1\over 2} \bar \epsilon^i \Gamma^m \psi_\mu \
,\nonumber\\
\delta\psi_{\mu i} &=& D_\mu (\hat \omega) \epsilon_i +  { i\over 6 } (e^{{\phi
\over 3}}
F_{\rho \sigma ij}- {1\over 2\sqrt 2} e^{-{2\over 3}\phi} G_{\rho \sigma}
\Omega_{ij})
\Bigl(\Gamma_\mu{}^{\rho \sigma} - 4 \delta_\mu{}^ \rho  \Gamma^\sigma \Bigr)
\epsilon^j +\cdots \ ,  \nonumber\\
\delta A_\mu ^{ij}  &=& -{1\over \sqrt 3} e^{-{\phi \over 3}}
\Bigl ( \bar \epsilon^{[i} \Gamma_\mu \chi ^{j]}  + {1\over 4} \Omega^{ij}
\bar  \epsilon^k
\Gamma_\mu \chi _{k} \Bigr) -i  e^{-{\phi \over 3}}
\Bigl ( \bar \epsilon^{[i} \psi_\mu  ^{j]}   + {1\over 4} \Omega^{ij}
\bar  \epsilon^k
\psi_{\mu k} \Bigr)\ , \nonumber\\
\delta B_\mu &=& {1\over \sqrt 6}  (e^{2 \phi \over 3} \Bigl( \bar \epsilon^i
\Gamma_\mu \chi _i - {i\sqrt 3 \over 2} \bar \epsilon^i \psi_{\mu i}\Bigr) \ ,
\nonumber\\
\delta \chi_i &= & -{i\over 2\sqrt 3}\Gamma^\mu \partial_\mu  \phi \epsilon_i +
{1\over
2\sqrt 3}\Bigl (e^{{\phi \over 3}} F_{\rho \sigma ij} + {1\over \sqrt 2}
(e^{-{2 \phi \over 3}} G_{\rho \sigma}
\Omega_{ij}\Bigr) \Gamma ^{ \rho \sigma} \epsilon ^j + \cdots \ , \nonumber\\
\delta\phi &=& {i \sqrt 3\over 2} \bar \epsilon^i  \chi_i \ .
\end{eqnarray}
Here $\Omega_{ij}$ is a symplectic matrix. The truncation of N=4 supergravity
to N=2 supergravity interacting with one vector multiplet is carried out  by
keeping
$\psi_{\mu i}, \chi_i$
with ($i=1,2 $) only and $A_\mu {}^{12} = -A_{\mu}{}^{34} \equiv  {1\over 2}
A_\mu, B_\mu , g_{\mu\nu}$ and $\phi$. By inspecting the supersymmetry
transformations of the
truncated fields it is not difficult to see that this is
a consistent truncation. The bosonic action becomes\footnote{This action is
equivalent upon  rescalings  to the action of N=2 supergravity interacting with
one  vector multiplet as presented in the Appendix of \cite{GT2}.}:
\begin{equation}
e^{-1} {\cal L} = -{1\over 2} R   - {1\over 4} e^{{2\over 3}\phi} F_{\mu\nu} F
^{\mu\nu } - {1\over 4} e^{-{4\over 3}\phi} G_{\mu\nu} G ^{\mu\nu }
-{1\over 6} ( \partial_{\mu} \phi) ^2   +{ e^{-1}   \over 4
\sqrt2}\epsilon^{\mu\nu\rho\sigma\lambda} F_{\mu\nu}
F_{\rho\sigma  } B_\lambda \ .
\end{equation}
The supersymmetry  transformations of the fermionic fields with vanishing
fermion are
\begin{eqnarray}
\delta\psi_{\mu } &=& D_\mu (\omega) \epsilon - { i\over 12 }
\Bigl(\Gamma_\mu{}^{\rho \sigma} - 4 \delta_\mu{}^ \rho  \Gamma^\sigma \Bigr)
(e^{{\phi \over
3}}
F_{\rho \sigma } - {1\over \sqrt 2} e^{-{2\over 3}\phi} G_{\rho \sigma})
\epsilon  \ ,\nonumber\\
\delta \chi&= & -{i\over 2\sqrt 3}\Gamma^\mu  \partial_\mu \phi \epsilon -
{1\over 4\sqrt
3} \Gamma ^{ \rho \sigma}\Bigl (e^{{\phi \over 3}} F_{\rho \sigma} +  \sqrt 2
e^{-{2 \phi
\over 3}} G_{\rho \sigma}
\Bigr)  \epsilon    \ ,
\end{eqnarray}
and we have omitted the symplectic index on the spinors (to avoid confusion
with the index on $\phi^i$ in the previous section).

Our action can be compared with the action used in \cite{SV} if we rewrite it
as follows
\begin{equation}
e^{-1} {\cal L} = {1\over 2} \left ( - R   - {1\over 4} e^{{2\over 3}\phi}
F'_{\mu\nu} F
^{' \mu\nu } - {1\over 4} e^{-{4\over 3}\phi}  G'_{\mu\nu}  G ^{ '\mu\nu }
-{1\over 3} ( \partial_{\mu} \phi) ^2   +{ e^{-1}   \over 8
}\epsilon^{\mu\nu\rho\sigma\lambda} F'_{\mu\nu}
F'_{\rho\sigma  } B'_\lambda \right) \ ,
\end{equation}
where
\begin{equation}
F' = \sqrt 2 F\ , \qquad G' = \sqrt 2 G \ , \qquad B' = \sqrt 2 B \ .
\end{equation}
We observe a discrepancy in the kinetic term for the scalar field which is
however harmless since the solution has a constant moduli field.  Thus the
extreme black hole of Strominger-Vafa can be embedded into the N=2 supergravity
interacting with one constant vector multiplet.

To make contact with the formalism of previous section dealing with the general
case of the very special geometry, we make the following identifications: $i=1,
I=0,1$ and
\begin{eqnarray}
i=1, \qquad I=0,1 \ , \qquad A_\mu{}^1&\equiv& A_\mu \ , \qquad A_\mu{}^0
\equiv B_\mu \ ,   \nonumber\\
G_{11} = e^{{2\phi \over 3}}\ , \qquad G_{00} = e^{-{4\phi \over 3}} \ , \qquad
C_{011} &=&  2\sqrt 2 \ , \qquad g_{11}=1 \ , \qquad \phi^1 = {1\over \sqrt 3}
\phi \nonumber \ ,\\
q_1= - {8 \sqrt 2 Q_F\over \pi}  \ , \qquad q_0 &=& 2 \sqrt 2 Q_H \ .
\end{eqnarray}

Comparing the supersymmetry transformations we must identify:
\begin{equation}
{\partial t_1\over  \partial \phi^1} = - {1\over \sqrt 3} \left({4\over
3}\right) ^{2/3} e^{{\phi \over 3}} \ , \qquad
{\partial t_0 \over \partial \phi^1} = -  \sqrt {2 \over 3} \left( {4\over
3}\right) ^{2/3}e^{-{2\phi \over 3}} \ .
\end{equation}
To satisfy the relations of special geometry and in particular to have
$(t^I)_{,i} t_I=0$ we get
\begin{eqnarray}
t_1 &=&  -\left({4\over 3}\right) ^{2/3} e^{{\phi \over 3}} \ , \hskip 4 cm
t^1
=  {2\over 3}
 \left({3\over 4}\right) ^{2/3} e^{-{\phi \over 3}}\ , \nonumber\\
t_0 &=& {1\over \sqrt 2}  \left({4\over 3}\right) ^{2/3} e^{-{2\phi \over 3}} \
, \hskip 3 cm t^0 = {\sqrt 2\over 3}
 \left({3\over 4}\right) ^{2/3} e^{{2\phi \over 3}} \ .
\end{eqnarray}
The enhancement of supersymmetry near the horizon is provided by ${\partial t_I
\over \partial \phi} G^{IJ}q_J=0$ and we get the fixed value of the moduli near
the horizon:
\begin{equation}
e^{\phi} = - {1\over \sqrt2} {q_1\over q_0}\ , \qquad t^1 q_1 = {1 \over
2^{1/6}3^{1/3} }  \left ( q_0 (q_1) ^2 \right )^{1/3}, \qquad t^0 q_0 = {1
\over
2} t^1 q_1 \ .
\end{equation}
Now we can express the combination $t^Iq_I$ near the horizon required for the
entropy as
 \begin{equation}
t^Iq_I= {3^{2/3}\over 2^{7/6}}  \left (q_0 (q_1) ^2 \right )^{1/3}
\end{equation}
and
\begin{equation}
r_0^6 ={1\over 6^2} \{ (t^Iq_I)_{\rm hor} \}^3 \qquad \Longrightarrow  \qquad
{1\over 2^{11/2}} (-q_0)^2 q_1 = {8 ( Q_H Q_{F})^2  \over \pi^2}\ ,
\end{equation}
and
\begin{equation}
A= 2\pi^2 (r_0)^3 = {\pi^2\over 2}  \left(  {q_0\over \sqrt 2}\left[
{q_1\over \sqrt 2} \right]^2
 \right )^{1/2}.
\end{equation}

Thus  we have reproduced the Strominger-Vafa area formula as an example of our
general
area formula
\begin{equation}
A= 2\pi^2 (r_0)^3 = {\pi^2\over 3} \left\{ (C^{IJ} q_I q_J)_{\partial i
Z=0}\right\}^{3/4} \qquad  \Longrightarrow  \qquad  8\pi \sqrt {Q_H (Q_F)^2
\over 2} \ ,
\end{equation}
and therefore this black hole solution  fits
into  our consideration of the very special geometry.

For the general case of   $n $
N=4 vector multiplets with duality group $ O(1, 1) \times {O(5,n)\over
O(5)x)O(n)}$
the formula  for the largest eigenvalue of the central charge, extremized in
the moduli space was presented in \cite{FK2}. This translates into the N=4
duality invariant area formula
\begin{equation}
A=  8\pi \sqrt {Q_H (Q_F)^2
\over 2} \ ,
\label{4}\end{equation}
where $Q_H$ is the singlet charge and  $(Q_F)^2$ is the $O(5,n)$  Lorenzian
norm of the other $5+n$ charges $Q_F$ .

Upon   truncation  to $N=2$ theory this gives a very special geometry with
$C_{IJK}$
intersection of the form ($I,J= 0,1, \dots , n+1 )$ and ($i=1, \dots n+1)$
\begin{equation}
C_{IJK}=> \left (\matrix{
C_{0ij}= \eta_{ij}\cr
0 \; \; {\rm otherwise} \cr
}\right ) \ ,
\end{equation}
where  $\eta_{ij}$  is  the Lorenzian
metric of $O(1,n)$   and the very special geometry is $ O(1, 1) \times
{O(1,n)\over O(n)}$.  Upon truncation the area has the same form as in
(\ref{4})
\begin{equation}
A=  8\pi \sqrt {Q_H (Q_F)^2 \over 2}=8\pi \sqrt {Q_H (Q_i \eta_{ij} Q_j)\over
2}  \ ,
\end{equation}
where again $Q_H$ is the singlet charge and  $(Q_F)^2$ is the $O(1,n)$
Lorenzian norm of the other $1+n$ charges $Q_i$.

\section{ Discussion}

Thus we have given a complete description of 5d static black holes near the
horizon which can be embedded into N=2 and N=4 supergravity with arbitrary
number of vector multiplets.
They all show  enhancement of supersymmetry near the black hole horizon. In N=2
as well as in N=4  case the unbroken supersymmetry (1/2 in N=2 and 1/4 in N=4)
is doubled. As the result in all cases the dimension of the Killing spinor of
unbroken supersymmetry is the dimension of the Dirac spinor admitted by the
anti-deSitter space and the sphere.

The significance of the enhancement of supersymmetry near the horizon is
related to the fact
that in all cases when we have found finite horizon area of supersymmetric
solutions there was also the enhancement of supersymmetry present. This
concerns all four- and five-dimensional  static black holes where we deal with
the near-horizon geometries $AdS_{2} \times S^{2}$ and $AdS_{2} \times S^{3}$
respectively.  The area in both cases is given by the volume of the $S^2$ and
$S^3$ sphere.

In more general situations when the near horizon geometry is given by
$AdS_{p+2} \times S^{d-p-2}$  the area of the horizon is given by the volume of
the $S^{d-p-2}$ sphere times the volume of the torus of dimension $p$, which
for the supersymmetric solutions shrinks to zero near the horizon. As the
result, in the class of solutions with $AdS_{p+2} \times S^{d-p-2}$
near-horizon geometries one can not expect
finite area of the horizon per unit volume  except for black holes.  Examples
of such configurations with enhancement of  supersymmetry and still vanishing
area  of
the horizon include: $d=10, p=3$; $d=11, p=2, p=5$. This geometric observation
matches the recent analysis \cite{ADF} of the cases of the  vanishing entropy
 in which a duality invariant expression
in terms of integral charges does not exist. In all these situations
the extremum of the central charge does not occur at finite rational
values of the moduli.

Few more configurations with near-horizon geometry  $AdS_{p+2} \times
S^{d-p-2}$  are known \cite{GHT}. They include
$AdS_{3} \times S^{2}$  describing a 5d magnetic string and $AdS_{3} \times
S^{3}$  describing a 6d self-dual string.  It has not been established yet
whether they have enhancement of  supersymmetry near the horizon. A calculation
of the type which we have
performed here for 5d black holes  is required to identify the supersymmetric
Killing spinors with the geometric ones for the 5d and 6d strings.

{}From the perspective of this study it would be interesting to have a more
careful look into configurations with
$AdS_{2} \times S^{d-2}$ geometries near the horizon. These would have finite
non-vanishing area related to the volume of the $S^{d-2}$ sphere. Such
configurations
of the Reissner-Nordstrom-Tangherlini-type
are known to solve equations of motion of Einstein-Maxwell theory in any
dimension.\footnote{We are grateful to G. Horowitz for this observation.}
However in $d>5$ they do not seem to have a supersymmetric embedding, at least
such embeddings have not been found so far.

In conclusion, we have studied the mechanism of enhancement of unbroken
supersymmetry near 5d black hole horizon and we have found the entropy-area
formula for solutions of N=2 supergravity interacting with arbitrary number of
vector multiplets.

\section*{Acknowledgements}
 We are   grateful to  Gary Horowitz, Andrew
Strominger,  Amanda Peet,  Tom\'as Ort\'{\i}n, Arvind Rajaraman, Paul Townsend
 and Wing Kai Wong for stimulating discussions. The work of S. F. was supported
in part by DOE grant DE-FGO3-91ER40662
and by European Commission TMR programme ERBFMRX-CT96-0045.
G.W.G.  would
like to thank N. Straumann for hospitality and the Swiss National
Science Foundation for financial support at I.T.P, Zurich where part
of this work was carried out.
R. K.  was
supported by the
  NSF grant PHY-8612280 and by the Institute of Theoretical Physics (ETH) in
Zurich and CERN where part of
this work was done.

\vfill
\newpage


\newpage

\begin{thebibliography}{30}
\bibitem{G} G.W. Gibbons, Nucl. Phys. {\bf B207},  337 (1982).
\bibitem{KP} R.~Kallosh and A.~Peet, Phys.~Rev. {\bf D46},  5223 (1992);
hep-th/9209116.
\bibitem{GT} G. W. Gibbons and P. K. Townsend, Phys. Rev. Lett.
{\bf 71},  3754 (1993); hep-th/9307049.
\bibitem{K} R. Kallosh,
Phys. Lett. {\bf B282}, 80 (1992), hep-th/9201029.
\bibitem{GKLTT}G.W. Gibbons, D. Kastor, L.A.J. London, P.K. Townsend, and J.
Traschen,  Nucl. Phys.  {\bf B416}, 850 (1994); hep-th/9310118.
\bibitem{SV}  A. Strominger and C. Vafa, Phys. Lett. {\bf B379}, 99 (1996),
hep-th/9601029.
\bibitem{rot}  J. Beckenridge, R. Myers, A. Peet, and C. Vafa, ``D-Branes and
Spinning Black Holes,'' Harvard University preprint HUTP-96-A005 (1996),
hep-th/9602065.
\bibitem{TseyCvY} A.A. Tseytlin, Mod. Phys. Lett. {\bf A 11}, 689 (1996),
hep-th/9601177;
M. Cvetic and D. Youm, ``General Rotating Five Dimensional Black holes of
Toroidally Compactified Heterotic String," hep-th/9603100.
\bibitem{FK1} S. Ferrara and R. Kallosh, Phys. Rev. {\bf D 54}, 1514 (1996),
hep-th 9602136.
\bibitem{FK2} S. Ferrara and R. Kallosh, Phys. Rev. {\bf D 54}, 1525 (1996),
hep-th 9603090.
\bibitem{GST} M. G\"unaydin, G. Sierra, and P.  K.  Townsend, Nucl.  Phys.
{\bf B242}, 244 (1984); Nucl.  Phys.
{\bf B253}, 573 (1985).
\bibitem{dWvP} B. de Wit and A. van Proyen, Phys. Lett. {\bf 293}, 94 (1992).
\bibitem{Cadavid} A.C. Cadavid, A. Ceresole, R. D'Auria, and S. Ferrara, Phys.
Lett. {\bf B357}, 76 (1995), hep-th/9506144; G. Papadopoulos and  P.K. Townsed,
Phys. Lett. {\bf B357}, 300 (1995), hep-th/9506150;
 I. Antoniadis, S. Ferrara, and T.R. Taylor,
Nucl. Phys. {\bf B460}, 489 (1996), hep-th/9511108.
\bibitem{KSW} R. Kallosh, M. Shmakova and W.K. Wong, ``Freezing of Moduli by N
= 2 Dyons,''  hep-th/9607077.
\bibitem{Tangh} F. R. Tanghelini, Nuovo Cimento {\bf 27} 636 (1963);
R.C. Myers and M.J. Perry, Ann. Phys. (N.Y.) {\bf 172} 304 (1986).
\bibitem{GT2} G. W. Gibbons and P. K. Townsend, Phys. Lett. {\bf B356}, 472
(1995)
\bibitem{AwTow} M. Awada and P. K. Townsend, Nucl.  Phys.
{\bf B255}, 617 (1985).
\bibitem{GHT} G. W. Gibbons, G. Horowitz and P. K. Townsend, Class. Quant.
Grav. {\bf 12}, 297 (1995), hep-th/9410073.
\bibitem{ADF} L. Andrianopoli, R. D'Auria, and S. Ferrara,  ``General Extension
of Extended Supergravity in Diverse Dimensions,'' hep-th/9608015.
\end{thebibliography}
\end{document}